\documentclass[aps,prl,reprint,superscriptaddress,showpacs,preprintnumbers,amsmath,amssymb]{revtex4-1}
\usepackage{graphicx}
\graphicspath{{fig/}}
\usepackage{dcolumn}
\usepackage{bm}
\usepackage{color}
\begin{document}
\preprint{K. Saitoh et al.}
\title{Stress relaxation above and below the jamming transition}
\author{Kuniyasu Saitoh}
\email[]{kuniyasu.saitoh.c6@tohoku.ac.jp}
\affiliation{Research Alliance Center for Mathematical Sciences, Tohoku University, 2-1-1 Katahira, Aoba-ku, Sendai 980-8577, Japan}
\author{Takahiro Hatano}
\affiliation{Department of Earth and Space Science, Osaka University, 560-0043 Osaka, Japan}
\author{Atsushi Ikeda}
\affiliation{Graduate School of Arts and Sciences, University of Tokyo, Tokyo, 3-8-1, Japan}
\affiliation{Research Center for Complex Systems Biology, Universal Biology Institute, University of Tokyo, Komaba, Tokyo 153-8902, Japan}
\author{Brian P. Tighe}
\affiliation{Delft University of Technology, Process \& Energy Laboratory, Leeghwaterstraat 39, 2628 CB Delft, The Netherlands}
\date{\today}
\begin{abstract}
We numerically investigate stress relaxation in soft athermal disks to reveal critical slowing down when the system approaches the jamming point.
The exponents describing the divergence of the relaxation time differ dramatically depending on whether the transition is approached from the jammed or unjammed phase.
This contrasts sharply with conventional dynamic critical scaling scenarios, where a single exponent characterizes both sides.
We explain this surprising difference in terms of the vibrational density of states (vDOS), which is a key ingredient of linear viscoelastic theory.
The vDOS exhibits an extra slow mode that emerges below jamming, which we utilize to demonstrate the anomalous exponent below jamming.
\end{abstract}
%
\maketitle
%
Amorphous materials such as suspensions, emulsions, foams, and granular materials are important in engineering science,
and a better understanding of their rheology is of central importance in many manufacturing processes \cite{larson}.
Although the rheophysics of amorphous materials has a long history, predictive description of their elasto-visco-plastic response remains challenging \cite{review-rheol0}.
Recently, physicists have focused on steady shear flow near the jamming transition at the packing fraction $\phi_J$
\cite{rheol0,rheol7,rheol9,rheol12,pdf1,rheol5,rheol3,rheol6,rheol8,rheol10,rheol11,saitoh19}.
Sufficiently close to jamming ($\Delta\phi\equiv\phi-\phi_J\rightarrow\pm 0$),
rheological flow curves (stress-strain rate relations) can be collapsed to master curves \cite{rheol0},
reminiscent of critical scaling functions near a second order phase transition \cite{hohenberghalperin1969}.
Critical scaling implies the existence of a diverging time scale $t_{\rm sf} \sim |\Delta\phi|^{-\alpha}$.
While details of the jamming scaling scenario remain controversial,
there is general agreement that the exponent $\alpha$ is the same on either side of the transition, with estimates for its value ranging from $2$ to $3$ \cite{rheol13}.

Steady shear flow is not the only way to probe time scales near jamming -- viscoelastic tests can also be used.
Viscosities measured via stress relaxation or oscillatory shear need not match the steady state viscosity, though in practice they are often comparable \cite{larson}.
Unlike steady flow, numerical studies of viscoelasticity typically apply perturbations to an isotropic reference state.
This is an important distinction, because there is evidence that (nearly) jammed states encode their loading history in their vibrational spectrum.
Lerner et al.~related the divergence of $t_{\rm sf}$ to this self-organization under steady shear \cite{vm-unjam4},
and recently Ikeda et al.~identified similar effects under isotropic compression \cite{ikeda19}.
However, it remains unclear what happens when a system is prepared isotropically and then subjected to a transverse stress increment, i.e.~shear.

Above jamming, small amplitude oscillatory shear and stress relaxation tests in $D=2$ dimensions reveal a diverging relaxation time
$t_+^\ast \sim \Delta \phi^{-\nu_+}$ with $\nu_+ = 1$ \cite{rl2,rl3,rl4} -- the viscoelastic relaxation time also diverges, albeit more slowly than $t_{\rm sf}$.
This result is consistent with theoretical predictions for both $D = 2$ and $3$, which relate $t_+^\ast$ to a broadening distribution of overdamped eigenmodes near $\phi_J$ \cite{rl0}.
3D stress relaxation tests below jamming also reveal a diverging time scale, $t_-^\ast\sim|\Delta\phi|^{-\nu_-}$ \cite{hatano}.
However, the exponent $\nu_- \approx 3.3$ is much larger than unity, reminiscent of the exponent $\alpha$ for steady flow.
The large difference between $\nu_+$ and $\nu_-$ represents a surprising departure from conventional dynamic critical scaling,
where a single exponent describes the divergence on both sides of the transition \cite{hohenberghalperin1969}.
The conventional scenario is in fact observed in closely related systems of overdamped spring networks, with $\nu_+ = \nu_- = 1$ \cite{rl7}.
Hence from this perspective $\nu_-$ is the anomalous exponent.
Resolving this discrepancy and understanding its origin represent an important challenge in jamming and rheology.

In this Letter, we use molecular dynamics (MD) simulations of 2D stress relaxation tests to present the first simultaneous measurements of $\nu_+$ and $\nu_-$.
We verify the large difference between the two exponents, eliminating the possibility that their difference was spurious or due to different measurement techniques.
We further establish qualitative similarities between viscoelastic relaxation and steady state rheology below jamming, including the anomalously large value of $\nu_-$.
We relate this difference to the same self-organization process identified previously in steady flow and isotropic cooling.

\begin{figure}
\includegraphics[width=\columnwidth]{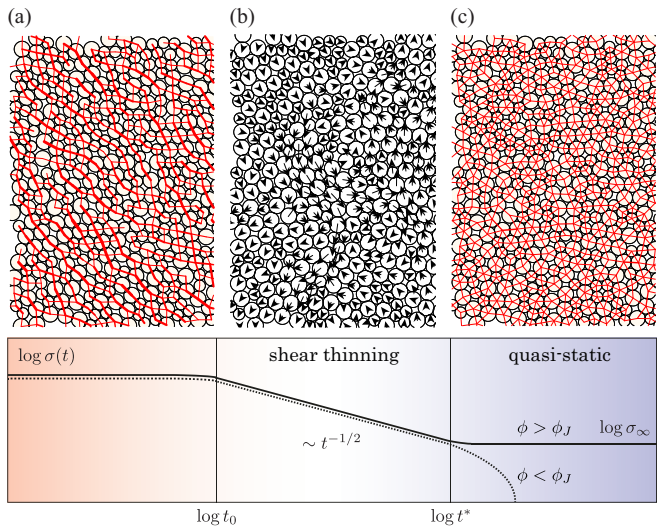}
\caption{
(a) Anisotropic force-chains (solid lines) immediately after applying a step strain $\gamma$, where the line width is proportional to the strength of elastic force between the disks (circles) in contact.
(b) Non-affine displacements $\delta\bm{u}_i$ (arrows) during the relaxation and (c) force-chains after the relaxation.
The lower panel is a sketch of \emph{stress relaxation} $\sigma(t)$, where the solid and dotted lines are the shear stress above and below the jamming transition, respectively.
The power-law relaxation $\sigma(t)\sim t^{-1/2}$ is observed in shear thinning regime $t_0<t<t^\ast$.
In quasi-static limit $t>t^\ast$, the shear stress is finite ($\sigma_\infty>0$) above jamming, while it drops to zero below jamming.
\label{fig:fch}}
\end{figure}
\emph{Numerical methods.}---
We perform simulations of $50:50$ binary mixtures of $N$ disks with diameter ratio $1.4:1$, a standard model in the jamming literature \cite{gn1,koeze16}.
Initial conditions are prepared  in a $L\times L$ square periodic box \cite{FIRE} at a controlled area fraction $\phi$ near the jamming point $\phi_J\simeq0.8433$.
The force between the disks, $i$ and $j$, in contact is modeled as $\bm{f}_{ij}^\mathrm{el}=k\xi_{ij}\bm{n}_{ij}$,
where $k$ is the stiffness and $\bm{n}_{ij}\equiv\bm{r}_{ij}/|\bm{r}_{ij}|$ with the relative position $\bm{r}_{ij}\equiv\bm{r}_i-\bm{r}_j$ is the normal unit vector.
The force is linear in the overlap $\xi_{ij}\equiv R_i+R_j-|\bm{r}_{ij}|>0$, where $R_i$ ($R_j$) is the radius of the disk $i$ ($j$).
The motion of the disk $i$ is described by overdamped dynamics \cite{rheol0,rheol7,pdf1,hatano},\
i.e.\ $k\sum_j\xi_{ij}\bm{n}_{ij}+\eta\dot{\bm{r}}_i=\bm{0}$, where $\eta$ is introduced as the damping coefficient.
In this model, the stiffness and damping coefficient determine a time scale $t_0\equiv\eta/k$.

\emph{Relaxation tests.}---
To study viscoelastic properties of a packing, we apply a small strain step $\gamma$ to the system,
where every disk's position $(x_i,y_i)$ is replaced with $(x_i+\gamma y_i,y_i)$ under the Lees-Edwards boundary conditions \cite{hatano}.
Figure \ref{fig:fch}(a) shows a snapshot of the anisotropic force-chains (solid lines) that develop immediately after the simple shear deformation.
The affine displacements $\bm{u}_i^\mathrm{affine}=(\gamma y_i,0)$ generate force imbalances,
and for $t > 0$ the particles are allowed to relax to a new mechanical equilibrium while $\gamma$ is held fixed via the Lees-Edwards boundary conditions.
As in Fig.\ \ref{fig:fch}(b), we observe complex non-affine displacements of the disks, $\delta\bm{u}_i=\bm{u}_i-\bm{u}_i^\mathrm{affine}$, during the relaxation.
After relaxation the initial force-chains are significantly weakened (Fig.\ \ref{fig:fch}(c)) and the shear stress has been reduced.
Shear stress is calculated as $\sigma(t)=L^{-2}\sum_{i>j}f_{ijx}^\mathrm{el}(t)r_{ijy}(t)$,
where $f_{ijx}^\mathrm{el}(t)$ and $r_{ijy}(t)$ are the $x$- and $y$-components of the elastic force and relative position, respectively.
We neglect infinitesimal kinetic contributions to the stress and subtract the \emph{pre-stress} (the shear stress in prepared packing) from numerical data to reduce noise.

\begin{figure}
\includegraphics[width=\columnwidth]{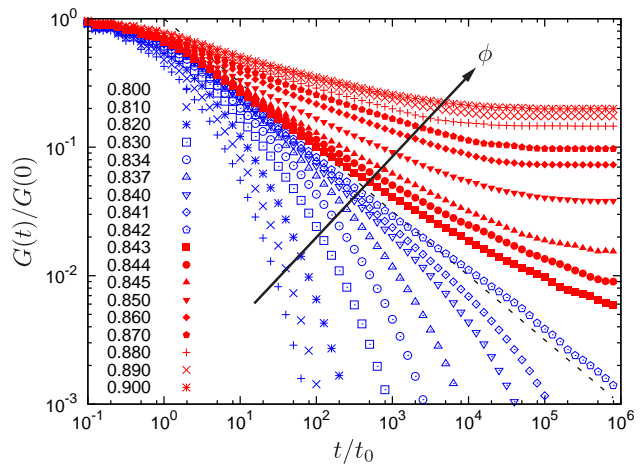}
\caption{
Double logarithmic plots of the scaled shear modulus $G(t)/G(0)$ and scaled time $t/t_0$, where the dotted line represents the power law decay $G(t)\sim t^{-1/2}$.
The area fraction $\phi$ increases as indicated by the arrow and listed in the legend,
where the (blue) open and (red) closed symbols represent the data below ($\phi<\phi_J$) and above jamming ($\phi>\phi_J$), respectively.
Here, $\gamma=10^{-2}$ is applied to $N=131072$ disks.
\label{fig:srelax}}
\end{figure}

If the strain step is sufficiently small, $\gamma\ll1$, the shear stress is described by linear viscoelasticity \cite{larson},
\begin{equation}
\sigma(t) = \int_{-\infty}^t G(t-t')\dot{\gamma}(t')dt' = \gamma G(t)~,
\label{eq:linearres}
\end{equation}
where $\dot{\gamma}(t)=\gamma\delta(t)$ is the imposed shear rate and $G(t)$ is the time-dependent shear modulus.
Figure \ref{fig:srelax} displays our numerical results of scaled shear modulus $G(t)/G(0)$, where the area fraction $\phi$ increases across the jamming point $\phi_J$ (as indicated by the arrow).
We observe instantaneous responses on a short time scale $t<t_0$ regardless of the area fraction.
Below jamming ($\phi<\phi_J$), the shear stress finally decays to zero, while it asymptotically decreases to a finite \emph{remnant stress} $\sigma_\infty=\gamma G_\infty$ above jamming ($\phi>\phi_J$).
As the system approaches the jamming transition, we find power-law decay $G(t)\sim t^{-1/2}$ in the intermediate time scale $t_0<t<t^\ast$,
where the relaxation time $t^\ast$ strongly depends on the proximity to jamming.
From these observations, we divide the time development of shear modulus into three different regimes:
(i) \emph{Instantaneous response} $G(t)\sim\mathrm{const.}$, (ii) \emph{shear thinning} $G(t)\sim t^{-1/2}$,
and (iii) \emph{quasi-static limit} $G(t)\sim 0$ for $\phi<\phi_J$ and $G(t)\sim G_\infty$ for $\phi>\phi_J$ \cite{rl0}.
%

\begin{figure}
\includegraphics[width=\columnwidth]{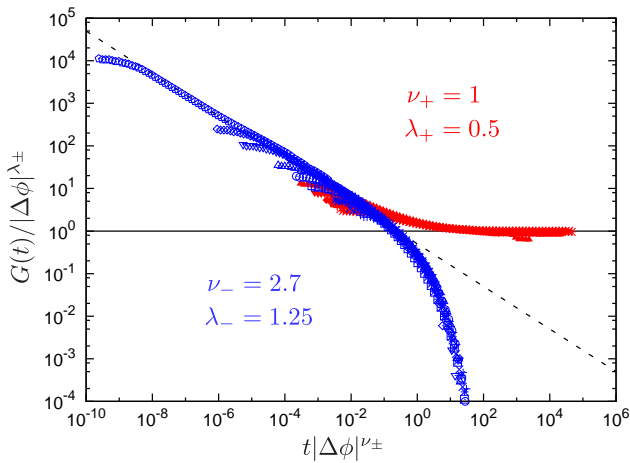}
\caption{
Data collapses of the time dependent shear modulus $G(t)$, where the dotted line and symbols are as in Fig.\ \ref{fig:srelax}.
The horizontal solid line indicates the static shear modulus $G_\infty\sim|\Delta\phi|^{\lambda_+}$.
See the text for the exponents, $\nu_\pm$ and $\lambda_\pm$.
\label{fig:collap}}
\end{figure}

\emph{Critical divergence of the relaxation time.}---
To estimate the relaxation time $t^\ast$, we collapse the data of the shear modulus.
Figure \ref{fig:collap} plots $G(t)$ against time, where they are scaled by $|\Delta\phi|^{\lambda_\pm}$ and $|\Delta\phi|^{\nu_\pm}$, respectively.
The data above jamming (closed symbols) are well collapsed by the exponents, $\lambda_+=0.5$ and $\nu_+=1.0$,
implying emergence of the static shear modulus $G_\infty\propto\Delta\phi^{1/2}$, and divergence of the relaxation time $t^\ast\sim\Delta\phi^{-1}$
\footnote{A smaller strain step $\gamma=10^{-4}$ is used for the data above jamming in Fig.\ \ref{fig:collap} to demonstrate the linear responses \cite{rl3,ys2}.}.
These results are consistent with prior numerical measurements \cite{rl2,rl3,rl4} and theoretical predictions \cite{rl0}.
On the other hand, we find excellent data collapse for data below jamming (open symbols) using $\lambda_-=1.25$ and $\nu_-=2.7$.
Note that the {scaling relation} $\lambda_\pm/\nu_\pm\simeq 1/2$ is satisfied both above and below jamming (see Supplemental Material (SM) \cite{SupplMater}).
Therefore, the relaxation time  scales differently on either side of jamming,
\begin{equation}
t^\ast\sim
\begin{cases}
\Delta\phi^{-1} & (\phi>\phi_J) \\
|\Delta\phi|^{-2.7} & (\phi<\phi_J)
\end{cases}~.
\label{eq:tau}
\end{equation}
%

\emph{Linear viscoelastic theory.}---
We now rationalize the two different scaling relations of Eq.\ (\ref{eq:tau}) using viscoelastic theory \cite{rl0} and numerical measurements of the relaxation spectrum.
To begin, it is useful to note certain properties of $\imath \omega G^*(\omega)$, the Fourier transform of $G(t)$.
Here $G^\ast = G'+\imath G''$ is known as the complex shear modulus, while $G'$ and $G''$ are the storage and loss moduli, respectively.
Because $G(t)$ is a real-valued function, $G'$ ($G''$) must be even (odd) in $\omega$.
Assuming $G^\ast$ is analytic and taking the limit $\omega\rightarrow 0$, it follows that $G'-G_\infty\sim\omega^2$ and  $G''\sim\omega$.
(We recall that $G_\infty\equiv G'(0)$ is zero in a fluid and finite in a solid.)
The relaxation time is the scale where these elastic and viscous contributions to the stress balance:
\begin{equation}
t^\ast = \lim_{\omega\rightarrow 0}\frac{G'-G_\infty}{\omega G''} \,.
\label{eq:t_ast1}
\end{equation}

We will use Eq.~(\ref{eq:t_ast1}) to evaluate $t^\ast$ within the harmonic approximation by expanding about a reference state.
We choose the state at $t \rightarrow \infty$, as determined from numerics.
Above jamming, one could also choose the initial condition as a reference state because, for any finite sized system, the strain step can always be taken small enough that no contacts are made or broken.
Below jamming, however, the initial condition contains only ``kissing'' contacts and the initial, affine shear step necessarily alters the contact network, which violates the harmonic approximation.
At long times, however, the contact network assumes its final topology (see SM \cite{SupplMater}) and the approximation can be applied.
Given a reference configuration, the complex shear modulus can be calculated from
\begin{equation}
G^\ast(\omega) = \left\{G_a+\imath \eta\omega\right\} - L^{-D}\langle\tilde{\bm{u}}(\omega)|\left[\mathcal{K}-i\eta\omega\bm{1}\right]|\tilde{\bm{u}}(\omega)\rangle~.
\label{eq:G_ast}
\end{equation}
The ket $|\tilde {\bm u}(\omega)\rangle$ contains the Fourier transform of the particles' displacements from their reference positions, and $\mathcal K$ is the Hessian matrix.
The term $G_a$ is the affine shear modulus, which remains finite at jamming.
The second term on the righthand side of Eq.~(\ref{eq:G_ast}) is the non-affine contribution to the modulus.
A derivation of Eq.~(\ref{eq:G_ast}) is given in the SM \cite{SupplMater}.

Next we introduce the eigenfrequencies $\omega_n$ and eigenvectors $|n\rangle$~($n=1,\dots,2N$) of the Hessian $\mathcal{K}$.
By expressing the $2N$-dimensional vector $|\tilde{\bm{u}}(\omega) \rangle$ as a linear combination of the eigenvectors and taking the limit of vanishing frequency, Eq.~(\ref{eq:t_ast1}) becomes
\begin{equation}
t^\ast \sim \frac{\sum_n \Xi_n^2/\omega_n^6}{\sum_n \Xi_n^2/\omega_n^4} \,.
\label{eq:t_ast2}
\end{equation}
Here $|{\bm\Xi}\rangle$ is the net force imbalance per unit affine shear and $\Xi_n\equiv|\langle n|\bm\Xi\rangle|$ is its projection onto the $n^{\rm th}$ mode.
$\Xi_n$ is a measure of the mode's coupling to shear.

\begin{figure}
\includegraphics[width=\columnwidth]{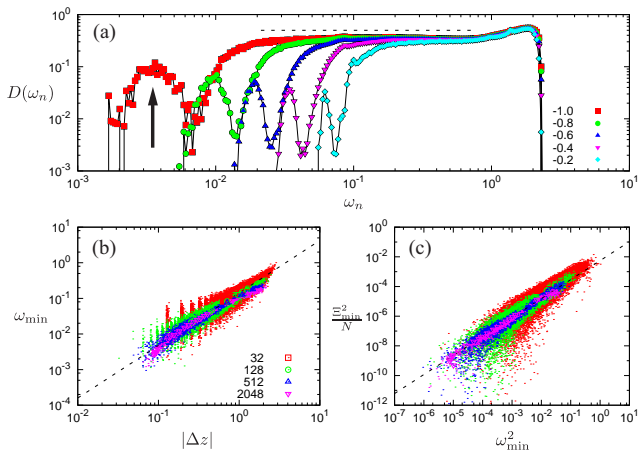}
\caption{
(a) Double logarithmic plots of the averaged vDOS for $N=2048$ disks below jamming, where $\log_{10}|\Delta z|$ increases as listed in the legend.
The dotted line and arrow indicate the plateau ($\omega^\ast<\omega_n$) and extra mode $\omega_\mathrm{min}$ for $|\Delta z|=10^{-1}$, respectively.
(b) and (c): Scatter plots of (b) $\omega_\mathrm{min}$ and $|\Delta z|$, and (c) $\Xi_\mathrm{min}^2/N$ and $\omega_\mathrm{min}^2$.
The open symbols are averages over each area fraction $\phi$, where the system size $N$ increases as listed in the legend of (b).
The dotted lines indicate (b) $\omega_\mathrm{min}\sim|\Delta z|^{1.47}$ and (c) $\Xi_\mathrm{min}^2\sim N\omega_\mathrm{min}^{2.56}$.
\label{fig:vdos}}
\end{figure}

\emph{The vibrational density of states.}---
In order to analyze the scaling of Eq.~(\ref{eq:t_ast2}) near jamming, it is necessary to quantify the spectrum of eigenfrequencies $\lbrace\omega_n\rbrace$.
We therefore introduce $D(\omega_n)$, the vibrational density of states (vDOS).
As shown previously \cite{vm0} and in the SM \cite{SupplMater}, the vDOS above jamming exhibits a plateau,
$D(\omega_n)\sim\mathrm{const.}$, extending to a characteristic scale $\omega^\ast$ that vanishes at unjamming \cite{gn3}.
We find a similar scenario holds below jamming: Fig.~\ref{fig:vdos}(a) reveals the same plateau over a range $\omega^\ast<\omega_n<t_0^{-1}$ (dotted line),
with $\omega^\ast\sim|\Delta z|^{1.3}$ controlled by the excess coordination number $\Delta z\equiv z-z_c$ ($z_c=4$ in two dimensions)
\footnote{The coordination number $z$ is measured after fully relaxing the system and removing \emph{floating disks} (with $z=0$).}.

In addition, as indicated by the arrow in Fig.\ \ref{fig:vdos}(a), there is a narrow band of slow modes below $\omega^\ast$.
(Note that packings below jamming also contain a number of \emph{floppy modes} with $\omega_n=0$, which are not visible in a log-log plot.)
The finite frequency band is centered on a peak value $\omega_\mathrm{min}$ that also goes to zero as the system approaches $\phi_J$ from below.
The finite width of this band is the result of averaging over $10^3$ different configurations
-- in fact, as we show in the SM \cite{SupplMater}, each packing contributes a single mode near $\omega_{\rm min}$, well separated from the modes at $\omega\gtrsim\omega^\ast$.
Such a ``special mode'' below jamming was previously observed by Lerner et al.~in steady shear flow \cite{vm-unjam4} and by Ikeda et al. \cite{ikeda19} in isotropic cooling.
At the same time, randomly cut under-coordinated spring networks have no such special mode, although their vDOS is otherwise similar \cite{vm-sprn0}.
These results suggest that loading trains the system to ``know about'' a particular direction in stress-space, by (self-)organizing to support the load.
This, in turn, is encoded in the special mode.
Our systems are prepared via an isotropic process and then subjected to  shear, so (unlike prior work) the stress increment is transverse to the preparatory load.
It is therefore not self-evident if or how the special mode should influence the dynamic viscosity.
%

\emph{Scaling analysis.}---
The exponent $\nu_+ = 1$ characterizing relaxation above jamming was derived in Ref.~\cite{rl0}, so we focus here on the case below jamming.
In order to perform scaling analysis, we numerically measure $\omega_\mathrm{min}$ and the coupling strengths.
Figure \ref{fig:vdos}(b) displays scatter plots of $\omega_\mathrm{min}$ and $|\Delta z|$, where each dot is the result of one configuration of the disks.
Each symbol is the average of $\omega_\mathrm{min}$ over packings with the same area fraction $\phi$.
The mode frequency scales as $\omega_\mathrm{min}\sim|\Delta z|^{1.47}$ (dotted line), independent of the system size $N$ (colors and symbols).

Excluding the special mode, the coupling strengths scale as $\Xi_n^2\sim\omega_n^2$, independent of $N$.
This relation, which implies that floppy modes have no shear coupling ($\Xi_n(\omega_n = 0) = 0$), has been validated numerically for {unstressed systems} above jamming \cite{vm4}.
In the SM \cite{SupplMater} we motivate it theoretically and confirm that it holds below jamming as well.
In sharp contrast, Fig.\ \ref{fig:vdos}(c) shows that coupling strengths for the special mode collapse
when plotting $\Xi_\mathrm{min}^2/N$ versus $\omega_\mathrm{min}^2$ for different sizes (colors and symbols).
This implies $\Xi_\mathrm{min}^2\sim N\omega_\mathrm{min}^{2.56}$ (dotted line).
Note, in particular, that $\Xi_\mathrm{min}^2$ is extensive, i.e.~it couples to shear differently, and more strongly, than other modes.
This strong coupling to shear is surprising, in the sense that the system was prepared isotropically.
Evidently, the reorganization triggered by a small shear step is sufficient to develop the special mode -- which does not occur above jamming.
We attribute this difference to the fact that any strain step can make or break contacts,
whereas finite size systems above jamming can be probed without inducing rearrangements \cite{rl2,rl8,cc0,cc1}.

Because the special mode is the softest non-floppy mode in the system and couples extensively to shear,
it dominates each of the sums on the righthand side of Eq.~(\ref{eq:t_ast2}), with all other modes contributing sub-dominantly.
One finds
\begin{equation}
t^\ast \sim \omega_\mathrm{min}^{-2} \sim |\Delta z|^{-2.94}
\end{equation}
which agrees with Eq.\ (\ref{eq:tau}) because $|\Delta z|\sim|\Delta\phi|$ \cite{SupplMater,nafsc2}.
Hence, while relaxation above jamming is controlled by $\omega^\ast$ \cite{rl0}, below jamming it is set by $\omega_{\rm min}$.
This explains the difference in scaling of the relaxation time on either side of the transition.

\emph{Summary and outlook.}---
In this study, we numerically studied stress relaxation in soft athermal disks.
The relaxation time diverges when the system approaches the jamming point -- but, unusually, the critical exponent depends on the direction of approach [Eq.\ (\ref{eq:tau})].
We calculated the vDOS and found its plateau above the characteristic scale $\omega^\ast$.
In contrast, the vDOS below jamming has a delta peak at a special mode $\omega_\mathrm{min}$ ($<\omega^\ast$).
Using scaling analysis, we showed that $\omega_{\rm min}$ controls long time relaxation below jamming by setting the value of $\nu_-$.
As the same approach correctly predicts $\nu_+$ above jamming, as well \cite{rl0}, we have unified the theoretical description of relaxation on either side of jamming.

The results presented here were in 2D. Recent work by Olsson has raised questions about the value of the scaling exponent for steady flow in 3D.
Hence there is a need for a careful future study of $t^*$ in 3D, building on the work of Hatano \cite{hatano}.
As the special mode has also been demonstrated in 3D \cite{vm-unjam4,ikeda19}, we expect our approach to apply in that case as well.
Other likely future directions include the role of the viscous force law \cite{rl4} and, in particular, the role of inertia,
where comparison with experiments \cite{longtimetail} is important for industrial applications.
%
\begin{acknowledgments}
We thank E. Lerner, W. Kob, F. Radjai, K. Miyazaki, T. Kawasaki, H. Hayakawa, J. Boschan, and K. Baumgarten for fruitful discussions.
This work was supported by KAKENHI Grant No.\ 16H04025 and No.\ 18K13464 from JSPS.
Some computations were performed at the Yukawa Institute Computer Facility, Kyoto, Japan.
BPT acknowledges support from the Dutch Organization for Scientific Research (NWO).
\end{acknowledgments}
%
\bibliography{viscoelastic}
\end{document}